\relax
\documentclass[letterpaper]{article} 
\usepackage{aaai22}  
\usepackage{times}  
\usepackage{helvet}  
\usepackage{courier}  
\usepackage[hyphens]{url}  
\usepackage{graphicx} 
\usepackage{natbib}  
\usepackage{caption} 
\DeclareCaptionStyle{ruled}{labelfont=normalfont,labelsep=colon,strut=off} 
\usepackage{algorithm}

\usepackage{adjustbox}
\usepackage{multirow}
\usepackage{multicol}
\usepackage{booktabs,colortbl}
\usepackage{amsmath}
\usepackage{algpseudocode}
\usepackage{subcaption}
\usepackage{blindtext}
\usepackage{dblfloatfix}
\usepackage[dvipsnames, x11names]{xcolor}
\usepackage[explicit]{titlesec}
\usepackage{lipsum}
\usepackage{tikz}
\usepackage{amsmath,amssymb}

\usepackage{xcolor}

\usepackage{array}
\usepackage{makecell}

\newcolumntype{L}[1]{>{\raggedright\arraybackslash}p{#1}}
\usepackage{newfloat}
\usepackage{listings}
\floatstyle{ruled}
\newfloat{listing}{tb}{lst}{}
\floatname{listing}{Listing}

\title{Social media algorithms can curb misinformation, but do they?}
\author {
    Chhandak Bagchi\textsuperscript{\rm 1},
    Filippo Menczer\textsuperscript{\rm 2},
    Jennifer Lundquist\textsuperscript{\rm 1},
    Monideepa Tarafdar\textsuperscript{\rm 1},\\
    Anthony Paik\textsuperscript{\rm 1},
    Przemyslaw A. Grabowicz\textsuperscript{\rm 1,\rm 3}\thanks{This research was conducted when Przemyslaw A. Grabowicz was a research assistant professor at the University of Massachusetts Amherst. Contact: grabowicz@cs.umass.edu.}
}
\affiliations {
    \textsuperscript{\rm 1} University of Massachusetts Amherst\\
    \textsuperscript{\rm 2} Indiana University\\
    \textsuperscript{\rm 3} University College Dublin\\
}

\begin{document}

\maketitle

A recent article in \textit{Science} by \citet{guess2023social} estimated the effect of Facebook’s news feed algorithm on exposure to misinformation and political information among Facebook users. However, its reporting and conclusions did not account for a series of temporary emergency changes to Facebook’s news feed algorithm in the wake of the 2020 U.S. presidential election that were designed to diminish the spread of voter-fraud misinformation \cite{ushouse2023social}. This issue may have led readers to misinterpret the results of that study and to conclude that the Facebook news feed algorithm used outside of the study period mitigates political misinformation as compared to reverse chronological feed.

From September 24th through December 23rd 2020, Guess et al. performed a randomized experiment measuring the effects of social media feeds on user behaviors and attitudes during the election campaign and its aftermath. The experiment provided a randomly assigned group of Facebook users with a news feed sorted in reverse chronological order. The effects of this intervention were then compared to the control condition – the Facebook news feed algorithm as it was implemented during the study period. The study stated in the abstract that “the chronological feed affected exposure to content: the amount of [...] untrustworthy content [users] saw increased,” concluding that “social media algorithms may not be the root cause of phenomena such as increasing political polarization.” Others interpreted it to mean that algorithms have little effect on exposure to problematic content \cite{budak2024misunderstanding}. These are crucial messages as we move into the 2024 U.S. presidential election season.

However, during the experiment, Meta introduced 63 “break-glass” changes to Facebook’s algorithmic news feed – not reported by Guess et al. – changing the control condition of their experiment. These temporary changes were designed to diminish the relative visibility of news content from untrustworthy sources immediately after the 2020 U.S. presidential election \cite{roose2020facebook, ushouse2023social}. While Guess et al. acknowledge that their results may have been different “if a different content ranking system were used as an alternative to the status quo feed-ranking algorithms,” the fact that the control condition changed during their experiment affects the validity and conclusion of their study.

\begin{figure*}[t]
\centering
\includegraphics[width=1.8\columnwidth]{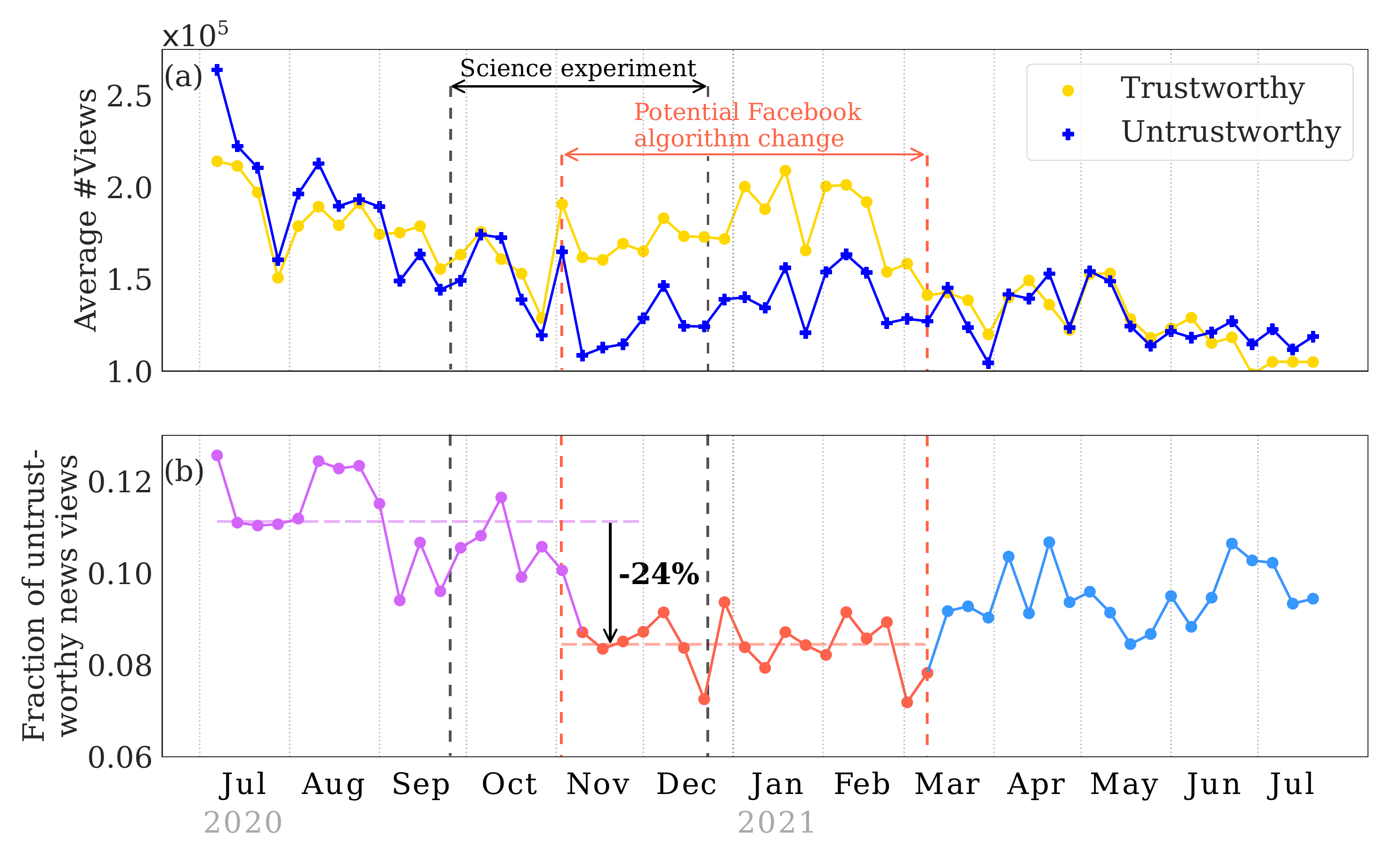}
\caption{ (a) Average weekly number of views of news from trustworthy and untrustworthy sources, calculated using the Facebook URLs dataset \cite{DVN/TDOAPG_2020}. Our estimates of untrustworthy news are based on links to sources rated mixed, low, or very low for factual reporting by Media Bias/Fact Check (MBFC) and shared at least 100 times, whereas Guess et al. consider any post by users with two or more reports as untrustworthy. (b) Fraction of views of untrustworthy news among all views. The horizontal dotted lines are averages of the points of the same color. We observe a drop during a period overlapping with the experiment, likely due to the changes in the news feed algorithm.
}
\label{fig:fig1}
\end{figure*}

The study by Guess et al. reports that the fraction of untrustworthy content was 40.9\% lower for the algorithmic news feed (their control) than for the reverse chronological news feed (their treatment). Using a different dataset provided by Meta, we measured the number of times users viewed news articles from trustworthy and untrustworthy news outlets in the year around the experiment. User exposure to news from trustworthy sources increased compared to untrustworthy sources from November 3, 2020, to March 8, 2021 \footnote{Our measurements are consistent with those of \citet{bandy2023facebook}. While they report a post-election drop in total referrals to both high- and low-quality news sources, we focus on the average views per news, which decreased for low-quality sources. They also report no substantial changes in the number of views of five prominent media outlets, while we analyze over a thousand sources.}. The period of potential impact of the news feed algorithm change (red arrow in Figure \ref{fig:fig1}a) coincides with the Guess et al. experiment (black arrow in Figure  \ref{fig:fig1}a). Such overlap might affect the reported effects of the chronological feed compared to the algorithmic news feed on the number of exposures to untrustworthy sources. Figure \ref{fig:fig1}b indicates that the fraction of untrustworthy news views decreased by around 24\% for the algorithmic feed. Therefore, a considerable portion of the decrease reported by Guess et al. may be attributable to the temporary algorithm changes. 

The implications of our estimated drop in the fraction of untrustworthy news views for the conclusions of Guess et al. depend on various factors. First, untrustworthy content can be measured in different ways (we find similar results when using NewsGuard scores instead of MBFC ratings). Second, the period of potential impact of the news feed algorithm change and the Guess et al. experiment period are not perfectly aligned. Finally, there may have been a surge in election-related misinformation \cite{vosoughi2018spread}, which might increase our estimated drop.

If there is a correlation between other variables in the Guess et al. study and exposure to untrustworthy sources, e.g., between the consumption of both misinformation and partisan news, then some other results presented by Guess et al. may also be explained by the temporary break-glass algorithm changes and may not replicate if the study were conducted again — a vivid demonstration of the importance of temporal validity when generalizing results from digital field experiments \cite{munger2019limited}. 

These results have research and societal implications. From a research perspective, they demonstrate the challenges of examining the effects of social media algorithms. The Guess et al. experiment was preregistered – that is, structured and declared ahead of its execution time. However, social media platforms do not register, let alone preregister, significant changes to their algorithms. For example, the exact effects and timing of most of the 63 break-glass news feed changes of 2020 are not known to the public. This can lead to situations where social media companies could conceivably change their algorithms to improve their public image if they know they are being studied. To prevent that, there is a need for independent research of social media platforms and consistent, transparent disclosures about major changes to their algorithms. Laws such as the Digital Services Act in the European Union and the proposed Platform Accountability and Transparency Act in the U.S., if properly enforced \cite{carvalho2024researcher}, could empower researchers to conduct independent audits of social media platforms and better understand the potentially serious effects of ever-changing social media algorithms on the public. In the last two years, however, social media transparency has diminished, e.g., on X and Reddit \cite{kupferschmidt2023twitter}, while Facebook has not updated the URLs dataset we analyzed here since 2022. Without access to such data, our understanding of the role of algorithms in curbing misinformation will remain incomplete. From a societal perspective,  our results suggest that news feed algorithms can mitigate misinformation. It might be valuable to keep the misinformation-preventing algorithmic feed in place even outside of election campaigns and their immediate aftermath. Unfortunately, there is no guarantee that algorithms beneficial to the public will be in place again in the future. Some Facebook employees claim in internal documents and interviews that the company ultimately chose to revoke the break-glass safeguards in the interest of market growth \cite{horwitz2023broken}. 

\section{Acknowledgements}
Authors acknowledge a data sharing agreement with Meta that enabled the study of Facebook's URLs dataset. However, Meta was not involved in this study in any way, financially nor intellectually, and did not oversee, nor review, the study. Ch.B., J.L., M.T., A.P., and P.A.G. acknowledge support by the Provost’s Interdisciplinary Research Grant 2023 (University of Massachusetts Amherst) titled ``Political Misinformation and Disinformation Through Social Media Biases''. F.M. is supported in part by the Knight Foundation and the Swiss National Science Foundation (grant 209250).
 
\bibliography{citations}

\end{document}